\documentclass[12pt,thmsa]{article}
\usepackage{sw20lart}

\input tcilatex
\QQQ{Language}{
American English
}

\begin{document}

\author{R.\ Peschanski\\CEA, Service de Physique
Theorique, CE-Saclay\\ F-91191 Gif-sur-Yvette Cedex, France}
\title{On dual descriptions of intermittency in a jet
\thanks{%
Written in honour of Andrzej Bialas for his 60th birthday and ever young
passion for Physics.}}
\maketitle

\begin{abstract}
Models of intermittent behaviour are usually formulated using a set of multiplicative
 random weights on a Cayley tree. However, intermittency in particle multiproduction from QCD jets 
 is related to fragmentation of   an   additive quantum number, e.g. energy-momentum.\
We exhibit the non-trivial stochastic mapping between these
 \textit{additive } and  \textit{%
multiplicative}  cascading processes.
\end{abstract}

\underline{1. Introduction: Intermittency and its formal description in
terms }

\underline{of mutiplicative cascading}

\medskip 

When Andrzej Bialas and I, 10 years ago, were puzzled by the mysterious multiplicity fluctuations
observed in an ultra-energetic cosmic ray event (JACEE collaboration), we
were far from realizing that a systematic study of multiparticle production
processes would emerge from  considering the  factorial moments of
the multiplicities distributions.\ The proposal we made$^{\text{\cite{bialas}%
}}$ was guided only by two seemingly reasonable requirements:

i) obtaining a non-subjective measure of fluctuations using the moments of the
multiplicity distribution,

ii) eliminating (as much as possible) the obvious source of statistical
fluctuations due to the finite number of particles produced by event.

We thus proposed the measurement of normalized factorial moments.
Under the (strong!) hypothesis that the statistical noise was simple, i.e. Poissonian in the
following formula (it was assumed Binomial in the case of the JACEE event because of the fixed multiplicity), one writes:

\begin{equation}
F_q(m)\equiv \frac{\left\langle n(n-1)...(n-q+1)\right\rangle_m}{%
\left\langle n\right\rangle _m^q}=\frac{\left\langle \rho ^q\right\rangle _m}{%
\left\langle \rho \right\rangle _m^q},  \tag{1}  \label{1}
\end{equation}

\noindent where $m$ labels a (suitably chosen) piece of available
phase-space for the reaction products, $n$ is the number of particles
registered in the bin for one event and $\rho $ is by definition an associated   continuous dynamical  variable corresponding to the local multiplicity density.

It is quite clear that the absence of fluctuations other than Poissonian
would give $F_q(m)=1.$ Moreover, varying the binsize $m$ allows one to
look deeper into short-distance fluctuations without being  hidden by the
 set of increasingly erratic statistical fluctuations. In the
same spirit, one would expect to observe dynamical objects, such
as clusters, resonances, jets -- by some signal
in the range of scales governed by the size of these objects.
However, Andrzej and I were soon confronted with a question of
interpretation: What means a continuous rise of the factorial moments with decreasing
phase-space size?  Indeed,  no obvious scale emerged from the observed rise  of 
factorial moments for  the JACEE event (this feature was later on confirmed by  systematic
studies of  various reactions).

At that time having no experience of such a
phenomenon in high-energy physics, we were inspired by the studies of turbulent  hydrodynamic flows$^{\text{\cite{mandelbrot,schertzer}}}$, in building a
mathematical toy-model  $%
(cf.$ $\alpha $-model, in the terminology and the version of Ref.$^{\text{\cite{schertzer}}}$ which we used after modification$^{\text{\cite{bialas}}}$) reproducing a pattern of multiplicity fluctuations
similar to what was surprisisingly suggested by the JACEE event.

 In its
simplest version, for one event, the model  is given by an
(infinite) set of randomly independent numbers $\left\{ W_s\right\} $ 
located
along the branches of a Cayley tree structure (see Fig.1a), where $s\ \left(
1\leq s\leq \nu \right) $ is a branching step while  $k_s\ \left( 1\leq k_s\leq
K\right) $ denotes which of the $K$ branches at a vertex bears the specific weight $W_s.$ The multiplicity
density profile is obtained for each individual bin by the following multiplication rule:

\begin{equation}
\rho_{\left[ m\right] }=\ \prod_1^\nu W_s\ ,  \tag{2}  \label{2}
\end{equation}

\noindent where the path $\left[ m\right] =\left\{ k_1,...,k_s,...k_\nu
\right\} $ is uniquely associated to the phase-space bin $m.$ Using the statistical independence of the weights and assuming many
 steps $\nu \gg 1,$ one gets:

\begin{equation}
\frac{\left\langle \rho _{\left[ m\right] }^{\ q}\right\rangle }{\left\langle
\rho _{\left[ m\right] }\right\rangle ^q}\ =\ \left\{ W^q\right\} _r^{\ \nu}\ 
\equiv\ \left(
M\right) ^{\frac{\ln \left\{ W^q\right\} _r}{\ln K}},  \tag{3}  \label{3}
\end{equation}

\noindent where $M (=K^\nu) $ is the total number of bins and, by definition:

\begin{equation}
\left\{ W^q\right\} _r =\ \int r^{.}(W)\ W^q\ dW    \tag{4}
\label{4}
\end{equation}

\noindent where $r(W)$ is a normalized weight probability distribution
satisfying $\left\{ 1\right\} _r =\ \left\{ W\right\} _r=1.$
As can be inferred from formulae (1-3), the multiplicity moments of
an $\alpha $-model (once the statistical noise is suitably
de-convoluted  ) show up whith a power-law dependence on the
number $M$ of phase-space bins, e.g. the resolution with wich one examines the system.\ This
behaviour is characteristic of the phenomenon of  intermittency in hydrodynamical turbulence; it  appears as a consequence of the random-cascading multiplicative property of the
 $\alpha $-model.

 Only a few years later,  the ubiquous
character of intermittent behaviour was recognised.\ In particular, an
a-priori different type of cascading behaviour, --though much more
familiar to particle physicists--, was suspected$^{\text{\cite{van hove}}}$
and then shown$^{\text{\cite{dokshitzer}}}$ to possess intermittency
properties. Interestingly enough, it is a direct consequence of Quantum
Chromodynamics --the field theory of strong interactions -- for the phase-space structure and development of quark and
gluon jets.\ In this framework, the cascading mechanism can be called 
{\it additive}, since at each elementary vertex energy-momentum is conserved and fragmented among the decay products along the cascade. This
is to be distinguished from  the $\alpha $-model, for which local densities are multiplied during the cascading process and thus not additively 
conserved .\ It is the subject of the present
paper to exhibit the transformation which asserts the equivalence
between {\it multiplicative} and {\it additive} cascading mechanisms and thus
their identical intermittency properties.

\medskip\ 
\eject
\underline{2. From a local to a global description of intermittent cascading}

\medskip\ 

As is explicited by the relation (3), the multiplicity  density moments
of the $\alpha $-model unravel the structure of fluctuations in the 
local limit  (i.e. short distance of order $1/M$). Our aim is now to
look for the system as a whole, i.e., its global description. For this sake it is convenient$^{\text{\cite{peschanski}}}$
to introduce the (random) Partition function $P_f(q)$ and an associated 
generating function $\mathcal{Z}_\nu(u);$ One writes

\begin{equation}
P_f(q)\equiv \frac 1M\sum_m\,\rho _{[m]}^q=\sum_{m=1}^M\,\prod_{s=1}\left( \frac{%
W_s}K\right) ^q;\ \mathcal{Z}_\nu (u)=\left\langle e^{-uP_f(1)}\right\rangle,
\tag{5}  \label{5}
\end{equation}

\noindent where one includes in the computation all the paths of the Cayley
tree for $\nu $ cascading steps, see Fig.1a.\ Note that a thermodynamical formalism can be usefully introduced$^{\text{\cite{peschanski}}}$ where $W_s/K$   acts like
a Boltzmann weight and $q$ as an inverse temperature.

Interestingly enough, $\mathcal{Z}_\nu (u)$ is known in Statistical Mechanics to obey a master equation$^{\text{\cite{derrida}}}.$
 Using an iterative
procedure ( $\nu  \rightarrow \nu +1$ 
)  adding one new step at the beginning of
the cascade and also the statistical independence of distinct sub-branching processes
(and after some work), one obtains:
\begin{equation}
\mathcal{Z}_{\nu +1}(u)=\left\{ \mathcal{Z}_\nu ^{\ K}\left( u\frac WK\right)
\right\} _r.  \tag{6}  \label{6}
\end{equation}

\noindent The compact formula (6) will be at the root of the
mathematical transformation between the  additive and multiplicative versions of
intermittent cascading. Let us for convenience introduce some
interesting extensions of  (6).

\smallskip\ 

\underline{Extension \# 1:} Inserting $P_f(q)$ instead of $P_f(1)$ in the generating
function (5), one gets a $q$-dependent master equation:

\begin{equation}
\mathcal{Z}_{\nu +1}(u,q)=\left\{ \mathcal{Z}_\nu ^{\ K}\left( u\left( \frac
WK\right) ^q, q \right) \right\} _r  \tag{6-1}  \label{6-1}
\end{equation}

\noindent which exhibits a scaling behaviour in terms of $q$
(or  tempera\-ture$^{\text{\cite{derrida}}}$.)\ Note that the equation
derived from (6-1) for the first moment reproduces the
intermittency property (3).

\smallskip\ 

\underline{Extension \# 2:} The master equation can be easily extended
to random-cascading processes including also random-branching (see Fig.1b). Introducing a time variable 
$t,$ an  $\epsilon $ probability of branching between $t$ and $t+\epsilon,
 $ a change of variables $\nu \rightarrow t\ ;\,\nu +1\rightarrow t +\epsilon ,$  and going to
the limit $\epsilon \rightarrow 0$ one gets

\begin{equation}
\frac{d\mathcal{Z}}{dt}\left( t;u\right) =\left\{ \mathcal{Z}^K\left( t;%
\frac{uW}K\right) \right\} _r-\mathcal{Z}\left( t;u\right)  \tag{6-2}
\label{6-2}
\end{equation}

\noindent Here too, the reader can check that the first moment equation
obtained by functional derivation with respect to $u$ leads to an
intermittency property (3) (with a modified exponent). Note that Eq. (6-2) takes the familiar form of a gain-loss formula.

\smallskip\ 

\underline{Extension \# 3:} Without modifying the basic
properties of the cascading process, one may  introduce a generalized
distribution $r (t;W_1,W_2,...,W_K).$
 Using a suitable change of variables, and after some
transformations of functions and variables one writes the following master
equation:

\begin{equation}
\frac{d\mathcal{Z}}{dt}(t;u)=\left\{ \prod^K_{i=1}\mathcal{Z}\left( t,u\frac{W_i}%
K\right) \right\} _r  - \mathcal{Z}(t;u).\tag{6-3}
\end{equation}
\noindent We shall soon recognize that equation (6-3) provides a generic form of the
additive cascading model of a jet based on QCD.

\medskip\ 

\underline{3. Additive vs. Multiplicative cascading models}

\medskip\ 

In particle physics theory, however, intermittent behaviour has not been
found directly under the form (6-1,3) of a multiplicative cascading process.\ It
appears in the study of the multiplicity of gluons and
quarks associated with an energetic jet in the framework of the resummed perturbative expansion of Quantum Chromodynamics. It has been for instance applied for the  decays of $\mathcal{Z}^{\circ } $s into quarks and gluons.
\ In the leading-logs approximation of perturbative QCD for 
jet calculations, one writes$^{\text{\cite{see}}}$ (for gluons):

\begin{equation}
\frac{\partial \mathcal{Z}}{\partial \ln Q}
\left( Q,u\right) =\frac 12\int
dz\ \Phi \left( Q,z\right) \ \left[ \mathcal{Z}\left( Qz,u\right) 
\mathcal{Z}\left( Q\left( 1-z\right) ,u\right) -\mathcal{Z}\left( Q,u\right)
\right]  \tag{7}  \label{8}
\end{equation}

\noindent where $\mathcal{Z}\left( Q,u\right) $ is the generating function
of gluon multiplicity factorial moments  from an initial gluon jet 
characterized by the virtuality $Q$; $\Phi \left( Q,z\right) $ is given in terms of the
renormalized QCD coupling constant $\alpha _s$ and of the triple gluon
Altarelli-Parisi-Kernel $P_{GG}^G$ (including quarks would transform (7)
into a two-by-two matrix form without changing our main conclusions).\ Note that
the important QCD property of {\it angular ordering} allows the derivation of  the
multiplicity distribution for any subjet  of initial energy $E$ and conical
aperture $\Theta $ starting from the same function $\mathcal{Z}\left( Q=E\Theta ,u\right) .$  As an important consequence, not only  global but also  local properties of multiplicity distributions of a QCD jet are determined by the solution of the equation (7). In particular angular ordering leads to the
property of angular intermittency.$^{\text{\cite{notion}}}$

As is explicit in formula (7), when compared  to formulae(6-1,3), the gluon cascading process is 
generated by the fragmentation of energy-momentum between gluons at the vertex,
 using the energy-fraction variable $z.$ Moreover, 
 identifying the ``time variable'' with  
$\ln Q,$ we find that Eq. (7) is \textit{not}\textbf{\ } defined with equal-time
observables (on contrary to Eqs. (6)), due to the mismatch between virtuality of
a gluon and energy sharing.

Yet, the two approaches are equivalent, as can be inferred$^{\text{\cite
{meunier1,meunier}}}$ from a crucial property of the multiplicity 
distributions, namely KNO scaling $%
^{\text{\cite{koba}}}$. This scaling property is verified in QCD$^{\text{\cite{see}}}
$ within the same conditions as Eq.\ (7). One may write, at least at high enough
virtuality\footnote{%
There could appear some problems near the boundaries of the $z$%
-integration in Eq.(7).
However  a check of    validity   can be made using
 Monte-Carlo simulations $^{\text{\cite{meunier2}}}$.}; the following
scaling relation:

\begin{equation}
\mathcal{Z}\left( Q,u\right) = \zeta\left( u\left\langle n\right\rangle 
_{\bf Q}\right)
,  \tag{8}  \label{9}
\end{equation}
where $\left\langle n\right\rangle_{\bf Q}$ is the average multiplicity
at virtuality $Q.$
\noindent Let us insert the KNO scaling relation (8) into (7). We get

\begin{eqnarray}
\frac{\partial \mathcal{Z}}{\partial \ln Q} &=&\frac12\int dz\ \Phi \left( Q,z\right) \times 
\nonumber   \\
&&  \tag{9}  \label{10} \\
&&\times \left[ \zeta\left( u\left\langle n\right\rangle _{\bf Qz}\right) \zeta\left(
u\left\langle n\right\rangle _{\bf Q(1-z)}\right)  -\zeta\left(
u\left\langle n\right\rangle _{\bf Q}\right) \right]   \nonumber
\end{eqnarray}

\noindent which, by a suitable change of variable can be cast into the
following equivalent form:
:

\begin{eqnarray}
\frac{\partial \mathcal{Z}\left( Q,u\right) }{\partial \ln Q} &=&\frac12\int\int
 dW_{1\;}dW_2\ \tilde r\left( Q;W_1,W_2\right)\times   \nonumber \\
&&  \tag{10}  \label{11} \\
&&\times\left\{ \mathcal{Z}\left( Q,u\frac {W_1}2\right) \mathcal{Z}\left( Q,u
\frac {W_2}2\right) -\mathcal{Z}\left( Q,u\right) \right\} ,  \nonumber
\end{eqnarray}

\noindent by defining:

\begin{equation}
W_1\equiv 2\ \frac{\left\langle n\right\rangle _{\bf Qz}}{\left\langle
n\right\rangle _{\bf Q}}\ ;\ W_2\equiv 2\ \frac{\left\langle n\right\rangle _{\bf Q\left(
1-z\right) }}{\left\langle n\right\rangle _{\bf Q}}  \tag{11}  \label{12}
\end{equation}

\noindent and
\begin{equation}
\tilde r \left( Q;W_1,W_2\right) \equiv \Phi \left( Q;z\right) \frac{dz(Q;W_1)}{dW_1}%
\delta\left( W_2-W_2(Q;W_1)\right) ,  \tag{12}  \label{13}
\end{equation}

\noindent where $ z\left( Q;W_1\right)$ and $W_2\left(Q;W_1\right)$
 are given in terms of the functional form of  $\left\langle n\right\rangle_{\bf Q} .$ This functional form  is obtained by solving
the linear equation coming from Eq.(7) for the  first moment, i.e. the first derivative with respect to $u.$ The solution of this equation is in general much simpler than for the generating function or higher moments. Note that a solution of this simpler equation  is sufficient to define the  transformation we look for (and thus to derive the intermittency properties).

Equations (10-12) are to be compared with  the generic equation (6-3). More precisely, the  change of function and variable  

\begin{eqnarray}
\left(\int\int
 dW_{1\;}dW_2\ \tilde r\left( Q;W_1,W_2\right)\right)^{-1}
\ \tilde r \left( Q;W_1,W_2\right)
&&\rightarrow \ r \left( Q;W_1,W_2\right)
  \nonumber \\
&&  \tag{13}  \label{14} \\
\left(\int\int
 dW_{1\;}dW_2\ \tilde r\left( Q;W_1,W_2\right)\right)
\ \partial \ln Q &&
\rightarrow \ \partial t,  \nonumber
\end{eqnarray}

\noindent gives the equivalence of  gluon cascading
 with a random branching, random cascading, {\it multiplicative} process. It amounts to choosing  in 
  Eq. (6-3) $K=2$ and the specific probability distribution $r,$ see (13).\ In a
sense, the \textit{multiplicative} formulation provides a statistical
description of the multiplicity density distribution at equal time (or
virtuality, or jet aperture), while the \textit{additive} one \textit{\ }%
describes the sharing of energy-momentum. Both descriptions are intimately
connected by the KNO scaling property. It is worth noticing that this specific stochastic process is, strictly speaking, of  {\it semi-random} type since
the value taken by the random weight $W_1$ fixes the other one 
$W_2$  at the same vertex.

\medskip\

\underline{4. Application to QCD gluon cascading}

\medskip\

Let us consider QCD cascading at the leading logarithmic approximation, by
keeping into account the effect of energy-momentum conservation at the
vertex. This problem has been recently raised  for the fluctuation pattern and solved$^{\text{\cite{meunier}}}$. In terms of \textit{%
global }observables, it amounts to solving equation (7) with the kernel $\Phi
\left( Q^2,z\right) \equiv 2\gamma _0^2/z$
where $\gamma _0^2\equiv 2\alpha_s N_c/%
\pi $ is kept fixed  for sake of simplicity.\ Using the formulation of Eqs. (10-13),
one gets:
\begin{eqnarray}
\frac{\partial \mathcal{Z}\left( Q,u\right) }{\partial \ln Q} &=&\gamma
_0\int \frac{dW_1}{W_1}  \times 
\nonumber \\
&&  \tag{14}  \label{15} \\
 &\times& \left[ \mathcal{Z}\left( Q,\frac{uW_1}2\right)\mathcal{Z}\left( Q,u\left[ 1-\left( \frac{W_1}2\right) ^{\gamma
_0^{-1}}\right] ^{\gamma _0}\right) - \mathcal{Z}\left( Q,u\right) \right] ,  \nonumber
\end{eqnarray}

\noindent where the relations: 

\begin{equation}
W_1\equiv \ 2 z^{\gamma _0}\;;\ W_2=2\ \left( 1-z\right) ^{\gamma _0},  \tag{15}
\label{16}
\end{equation}

\noindent are coming from the expression of the mean multiplicity $\left\langle
n\right\rangle _{\bf Q} \propto  Q^{\gamma _0};$  by inversion, one finds:

\begin{eqnarray}
z \!&\!\equiv&z\left( W_1\right) = \left(\frac {W_1}2\right) ^{1/\gamma
_0}\;;\ W_2  = 2\left[ 1-\left(\frac {W_1}2\right) ^{1/\gamma
_0}\right] ^{\gamma _0},  \nonumber \\
&&  \tag{16}  \label{17} \\
 & &\tilde r\left( Q;W_1,W_2\right)  \equiv\gamma _0\ \frac1{W_1}%
\ \delta \left( W_2- 2\left[ 1-\left(\frac {W_1}2\right) ^{1/\gamma_0}\right] ^{\gamma _0}\right) .
\nonumber
\end{eqnarray}

\noindent A few remarks are in order about the master equation (14).

 i) Comparing (14) with Eqn.(6-3) shows that $\mathcal{Z}$ is the
generating function for a random-branching process with triple vertex $%
1\rightarrow 2,$ (up to transformations like (13)). At each vertex, one branch corresponds to a randomly chosen
weight $W_1$ with the probability law $\gamma_0/W_1,$ see expressions (16). The second-branch weight is  then determined by
$W_2=2\left[ 1-\left( W_{1}/2\right) ^{1/\gamma _0}\right] ^{\gamma _0}.$
Interestingly enough, the local constraint of energy-momentum
conservation along the additive cascading process is transposed into a
specific stochastic law at the multiplicative vertex of the density 
cascading process.

 ii) In the approximation $W_1<2\;;\gamma _0\ll 1,$ then $\left( \frac
{W_1}2\right) ^{1/\gamma _0}\ll 1,$ and one can write a simplified version of
equation (13) namely:

\begin{equation}
\frac{\partial \ln \mathcal{Z}\left( Q,u\right) }{\partial \ln Q}=\gamma
_0\int \frac{dW_1}{W_1}\left( \mathcal{Z}\left( Q,\frac{uW_1}2\right)
-1\right) .  \tag{17}
  \label{18}
\end{equation}

\noindent Equation (17) has been found$^{\text{\cite{notion}}}$equivalent to the double-leading-$\log $ approximation of  QCD for the jet
 process, where one can neglect the recoil effect
upon the leading parton (quark or gluon). However, sizeable
values of $\gamma _0$ (which is still of order .5 at LEP energies) do
lead to substantial modifications of the multiplicity distributions$^{\text{%
\cite{see2}}}$ 
 and their  fluctuations in phase-space$^{\text{\cite
{meunier}}}$
due to energy-momentum conservation.
\medskip\ 

{\bf Acknowledgments}

I would like to express my admiration and friendship for Andrzej Bialas. During these last eleven years, it was  (and it will be in the future) a great pleasure to collaborate with him. I learned a lot from him, in particular  that listening to the differences and taking them into account provide the basis for a fruitful, equal-footing,  collaboration. 
I warmly thank  Jean-Louis Meunier for his  contributions to the topics contained in this paper and for the lively discussions about it. Thanks are due to Philippe Brax not only for his early contributions to the problem but also for his careful reading of the manuscript.

\bigskip

{\bf Figure Caption}

\bigskip
\textbf{a) Fixed branching}

Each path $[m]$ corresponds to a bin in phase space.\ The fluctuation density
in the bin is described by the product of random weights $W$ along the
path $[m].$ In the case of ``semi-randomness'' the decay is assymetric at
each step: one weight $W_1$ follows a random law, while $W_2$ is a function 
of  $W_1$ for the same vertex
($W_1$ or $W_2$ can be randomly left or right).\ The tree can be separated into
2 branches. Each of these branches defines in average $\mathcal{Z}_\nu$ of
the
Partition function, the whole  tree itself corresponding to $\mathcal{Z}_{\nu+1}.$

\textbf{b) Random branching}

Same structure as in Figure 1.a, but with small increments $\varepsilon $ of
the
number of steps $\nu$ and random-branching with uniform probability $%
\varepsilon .$
\eject

\input epsf

\epsfysize=18cm{\centerline{\epsfbox{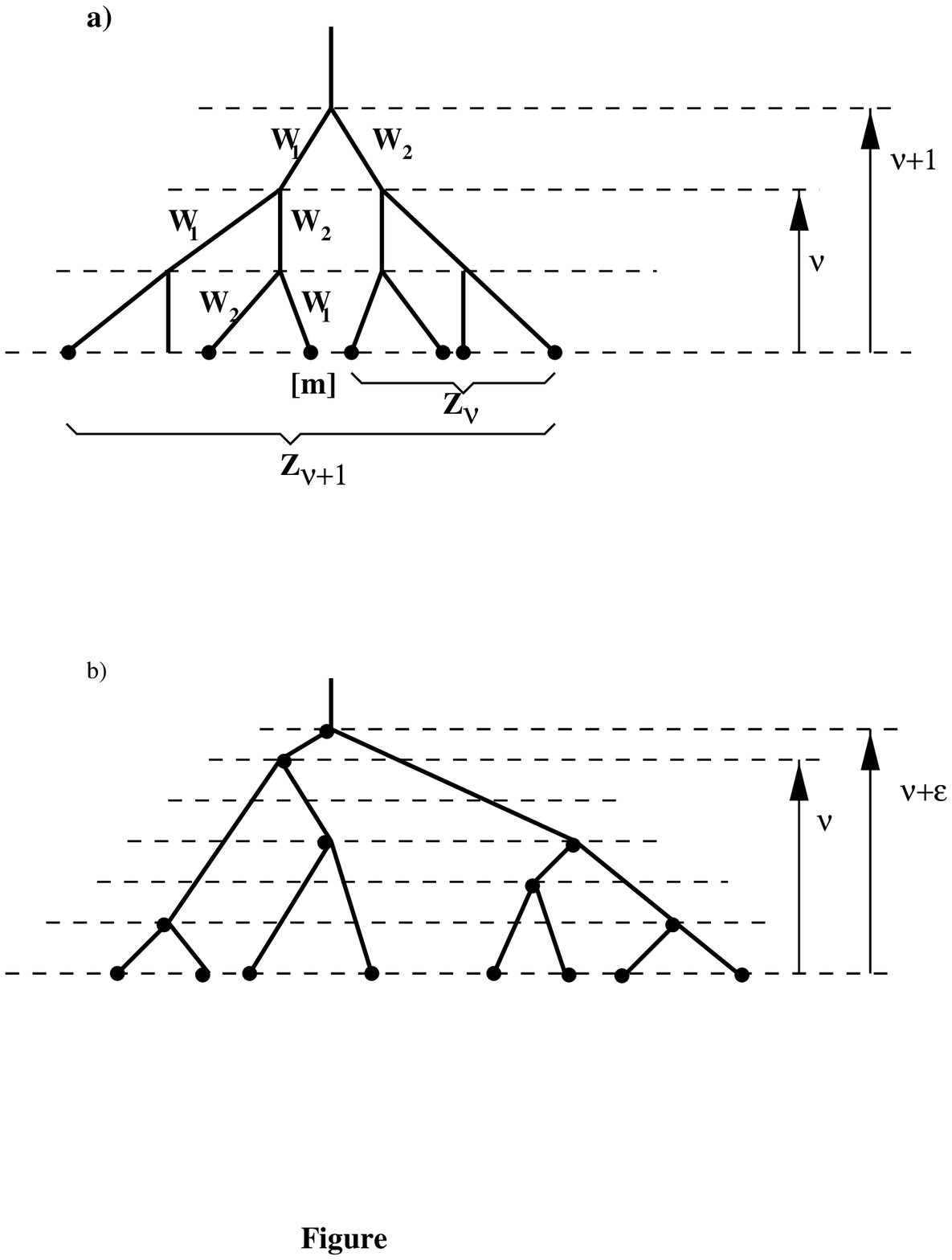}}}
\bigskip
\eject


\begin{thebibliography}{99}
\bibitem{bialas}  A. Bialas and R.\ Peschanski, \textit{Nucl. Phys. }\textbf{%
B273} (1986) 703, \textbf{B308} (1988) 857, \textit{Phys. Lett. }\textbf{%
B207 }(1988) 59.

\bibitem{mandelbrot}  B.B. Mandelbrot, \textit{J. Fluid.\ Mech.\ }\textbf{62}
(1974) 331; U.\ Frisch, P.L. Sulem and M.\ Nelkin, \textit{J.\ Fluid Mech. }%
\textbf{87} (1978) 718;

\bibitem{schertzer}  D. Schertzer and S. Lovejoy, in ``Turbulent Shear flows
4'' eds.\ L. J.S. Bradbury et al. (Springer, 1984).

\bibitem{van hove}  L. Van Hove, private communication (1989);

\bibitem{dokshitzer}  Y. Dokshitzer and I. Dremin, \textit{Nucl.\ Phys.\ }%
\textbf{B402} (1993) 139; R. Peschanski, in ``Proceedings of the XXII
International Symposium on Multiparticle Dynamics'' Santiago de Compostela,
Spain, July 1992 (C. Pajares editor, World Scientific) P.\ 164; 
P.Brax,
J.-L.\ Meunier and R.\ Peschanski, \textit{Zeit.\ f\"{u}r\ Physik}\textbf{\
C62} (1994) 649;\ W.\ Ochs and I.\ Wosiek, \textit{Phys.\ Lett.\ }\textbf{%
B305} (1993) 144.

\bibitem{peschanski}  R. Peschanski, \textit{Nucl.\ Phys.\ }\textbf{B327}
(1989) 144; in a different but equivalent language, see: A. Bialas, A.
Szczerba and K. Zalewski, \textit{Zeit. f\"{u}r. Phys.\ }\textbf{C46 }(1990)
163.

\bibitem{derrida}  B. Derrida and H. Spohn, \textit{J. Stat.\ Phys.}\textbf{%
\ 51 }(1988) 817, and references therein.

\bibitem{see}  See, for instance, the book: ``Basics of Perturbative QCD'' 
Yu. L. Dokshitzer, V.A. Khoze, A.H. Mueller and S.I. Troyan (Editions
FRONTIERES, FRANCE), p. 124 - 169, and references therein.

\bibitem{notion}   see P. Brax et al. in  \cite{dokshitzer}.

\bibitem{meunier1}  I thank J.-L.\ Meunier for having first pointed out to me the  relation between the two formalisms of intermittent behaviour by the generating function.

\bibitem{meunier}  J. -L.\ Meunier and R. Peschanski, ``Energy Conservation
Constraints on Multiplicity Correlations in QCD jets'', hep-ph/9603294 to
appear in \textit{Zeit. f\"{u}r Phys. }{\bf C}.

\bibitem{koba}  Z. Koba, H.B. Nielsen and P.\ Olesen, \textit{Nucl.\ Phys.\ }%
\textbf{B40 }(1972) 317; On the r\^{o}le of KNO scaling in the $\alpha $%
-model see: Y. Gabellini, J.-L.\ Meunier and R. Peschanski, \textit{Zeit.\
f\"{u}r Phys. }\textbf{C55} (1992) 455.

\bibitem{meunier2}  J.-L.\ Meunier, private communication.

\bibitem{see2}  See, for instance, Yu.L.\ Dokshitzer, \textit{Phys.\ Lett. }%
\textbf{B305} (1993) 295. See also F. Cuypers, {\it Zeit. fur Phys.} {\bf C54}
(1992) 87.
\end{thebibliography}
\end{document}